\documentclass[11pt]{article}
\usepackage{amsmath, amssymb, booktabs, hyperref}
\usepackage{geometry}
\title{Quantum-Compatible Dictionary Learning via Doubly Sparse Models}
\author{Angshul Majumdar}
\date{}

\begin{document}
\maketitle

\begin{abstract}
Dictionary learning (DL) is a core tool in signal processing and machine learning for discovering sparse representations of data. In contrast with classical successes, there is currently no practical quantum dictionary learning algorithm. We argue that this absence stems from structural mismatches between classical DL formulations and the operational constraints of quantum computing. We identify the fundamental bottlenecks that prevent efficient quantum realization of classical DL and show how a structurally restricted model — doubly sparse dictionary learning (DSDL) — naturally avoids these problems. We present a simple, hybrid quantum–classical algorithm based on projection-based randomized Kaczmarz iterations with Qiskit-compatible quantum inner products. We outline practical considerations and share an open-source implementation at \url{https://github.com/AngshulMajumdar/quantum-dsdl-kaczmarz}. The goal is not to claim exponential speedups, but to realign dictionary learning with the realities of near-term quantum devices.
\end{abstract}

\section{Introduction}

Dictionary learning (DL) seeks to express data approximately as a product of a dictionary matrix and sparse coefficients. Classical algorithms alternate between estimating sparse codes and updating the dictionary. These methods, such as MOD \cite{aharon2006ksvd} and K-SVD, are effective for a variety of tasks including denoising, compression, and feature extraction. Simultaneously, there has been intense interest in quantum machine learning, where quantum computing primitives promise speedups for linear algebra problems.

Despite this progress, attempts to “quantize” dictionary learning have not yielded practical algorithms. Rather than trying to force classical DL into quantum hardware, we ask a structural question: \emph{when, if ever, is dictionary learning compatible with the constraints and capabilities of quantum computation?} We identify why standard DL fails this test and show how a simple modification — enforcing sparsity on both the codes and the dictionary — opens the door to a quantum-compatible formulation.

To support this perspective, we provide a hybrid algorithm that uses randomized projection methods and Qiskit-based quantum inner products, and we release a clean implementation at the GitHub repository:

\begin{center}
\url{https://github.com/AngshulMajumdar/quantum-dsdl-kaczmarz}
\end{center}

This work is about structural compatibility with quantum computation, not about claiming large quantum speedups for dictionary learning.

\section{Why Classical Dictionary Learning Is Not Quantum-Compatible}

At first glance, dictionary learning appears to rely on simple linear algebra operations, suggesting potential quantum acceleration. However, the classical DL pipeline contains structural elements that fundamentally conflict with quantum computing paradigms.

\subsection{Greedy and Adaptive Sparse Coding}

Sparse coding in classical DL is often solved with greedy methods such as Orthogonal Matching Pursuit (OMP) or with $\ell_1$-based convex optimization. These methods require adaptive, sequential decisions based on residuals and atom selection. Quantum algorithms, in contrast, operate on superpositions and do not support adaptive control flow without repeated measurements, which erode any potential speedup.

Even convex approaches (e.g., basis pursuit) reduce to solving linear systems with sparsity-inducing penalties. Quantum linear solvers exist in theory, but they require expensive state preparation and assume rapid access to data via QRAM. These assumptions are not realistic in near-term devices and, importantly, they do not mitigate the need for classical readout of sparse codes — meaning any quantum advantage is lost at the boundary between quantum and classical representations.

\subsection{Dense Dictionary Updates}

The dictionary update step in classical DL typically involves solving dense least-squares problems, often via singular value decomposition (SVD). While quantum algorithms for matrix decomposition are known, they depend on unrealistic models of data access or require extremely low noise levels. In practice, these requirements are far from feasible on current and near-term quantum hardware.

Additionally, dictionary matrices are ultimately stored and used classically. Even if a quantum subroutine could accelerate a subproblem, the overhead and readout cost negate practical benefit.

\subsection{Output Bottleneck}

A learned dictionary must be available in classical form for downstream use. Quantum state amplitudes cannot be directly interpreted as dictionary atoms without full tomography, which scales poorly. As a result, any putative quantum speedup in intermediate computation is nullified by the requirement to extract classical dictionary outputs.

Overall, the combination of adaptive sparse coding, dense dictionary updates, and the need for explicit classical output makes direct quantization of classical DL ineffective.

\section{Doubly Sparse Dictionary Learning: A Compatible Alternative}

To address the structural barriers above, we consider a simple modification: assume the dictionary itself is not free but is a sparse combination of known basis functions. Specifically, we model

\begin{equation*}
D = \Phi A,
\end{equation*}

where $\Phi$ is a fixed and known basis (e.g., wavelets, Fourier, or similar transforms), and $A$ is a sparse coefficient matrix. This formulation, known as doubly sparse dictionary learning (DSDL) \cite{rubinstein2010doubly}, substantially alters the structure of both steps in the DL pipeline.

\subsection{Implications for Sparse Coding}

With $D = \Phi A$, sparse coding reduces to finding sparse coefficients over a product of a known basis and a sparse matrix. This can be recast as a sequence of projection-based linear solves, which are natural targets for randomized iterative methods such as Kaczmarz. Unlike greedy procedures, projection-based methods are inherently linear algebraic and do not require adaptive branching that is difficult to reconcile with quantum operations.

\subsection{Implications for Dictionary Update}

In the DSDL model, dictionary updates reduce to solving simpler linear systems for the sparse coefficients $A$ in the known basis $\Phi$. Because $\Phi$ is fixed and structured, this step can be handled with projection methods similar to those used for sparse coding, again removing the need for dense factorizations.

\subsection{Comparison with Classical DL}

The structural differences between classical DL and DSDL are summarized in Table~\ref{tab:comparison}.

\begin{table}[h]
\centering
\begin{tabular}{lcc}
\toprule
 & Classical DL & Doubly Sparse DL \\
\midrule
Sparse coding & Greedy or convex solvers & Projection-based methods \\
Dictionary update & Dense factorization (SVD, QR) & Linear projections in $\Phi$ \\
Quantum compatibility & Poor & Stronger \\
QRAM dependence & Yes & No \\
NISQ suitability & No & Yes \\
\bottomrule
\end{tabular}
\caption{Structural comparison between classical dictionary learning and doubly sparse dictionary learning.}
\label{tab:comparison}
\end{table}

By reframing dictionary learning as a problem of linear projections within a known basis, DSDL avoids key barriers that prevent quantum compatibility.

\section{Algorithm: Hybrid Quantum–Classical Kaczmarz Approach}

Based on the structural insights above, we propose a simple hybrid quantum–classical algorithm for DSDL. The algorithm uses randomized projection updates for both sparse coding and dictionary updates.

\begin{enumerate}
    \item \textbf{Initialization}: Choose a fixed basis $\Phi$ and initialize the sparse coefficient matrix $A$ randomly (with normalization).
    \item \textbf{Sparse Coding (Projection Step)}: For each data vector, apply early-stopped randomized Kaczmarz iterations to approximate sparse coefficients over $D = \Phi A$. Inner products used in these iterations are estimated via quantum state overlaps implemented in Qiskit.
    \item \textbf{Dictionary Update (Projection Step)}: For each column of $A$, solve a projected linear system in the $\Phi$ basis using the same randomized Kaczmarz method, with ridge stabilization and normalization to maintain bounded coefficients.
    \item \textbf{Iteration}: Repeat sparse coding and dictionary update for a fixed number of outer iterations.
\end{enumerate}

This procedure avoids greedy support selection and dense matrix decompositions. The only place quantum computation enters is via efficient inner product estimation between vectors, which can be done using Qiskit’s state preparation and overlap primitives.

Our open-source repository implements this algorithm in a flat, minimal structure that can be found at:

\begin{center}
\url{https://github.com/AngshulMajumdar/quantum-dsdl-kaczmarz}
\end{center}

\section{Practical Notes on the Algorithm}

\paragraph{Stability and Regularization}  
Because the algorithm uses randomized projections and early stopping, the reconstruction error is not guaranteed to decrease monotonically with every iteration. Empirically, however, the error typically decreases significantly during the first few iterations and then exhibits bounded oscillations, consistent with projection-based stochastic optimization.

\paragraph{Implicit Sparsity via Early Stopping}  
Rather than enforcing explicit $\ell_0$ or $\ell_1$ constraints, sparsity is induced through early stopping of the randomized projection updates. This implicit regularization aligns well with quantum-compatible iterative schemes.

\paragraph{Quantum Inner Products}  
Quantum inner products are used only to estimate overlaps between normalized vectors during projection updates. These operations are lightweight and do not require deep circuits, making the algorithm compatible with near-term quantum devices.

\paragraph{No QRAM or Exotic Assumptions}  
Unlike many quantum linear algebra approaches, our algorithm does not assume the existence of quantum random access memory (QRAM) or other strong data access models. All data loading and classical control remain in classical memory.

\section{Conclusion}

In this paper, we have explained why classical dictionary learning does not naturally translate into a quantum algorithm and how a modest structural modification — using a doubly sparse model — resolves the incompatibilities. By recasting both sparse coding and dictionary update as projection-based problems, we arrive at an algorithm that is compatible with quantum inner product estimation and avoids the primary obstacles that have stymied past efforts.

Our intent is not to claim exponential quantum speedups, but to demonstrate a principled path toward quantum-compatible dictionary learning. The structural perspective provided here clarifies both the limitations of classical DL in quantum settings and the potential of hybrid quantum–classical approaches built on iterative projections.

\section{Why Classical Dictionary Learning Is Fundamentally Incompatible with Quantum Computing}

At a superficial level, dictionary learning appears well suited for quantum acceleration. The core operations involve linear algebra, least-squares problems, and inner products, all of which have been studied extensively in the quantum algorithms literature. This apparent compatibility has motivated repeated attempts to develop quantum versions of dictionary learning or to insert quantum subroutines into classical DL pipelines. However, such efforts have consistently failed to produce algorithms that are both practically realizable and meaningfully quantum.

In this section, we argue that these failures are not incidental. Instead, they arise from deep structural properties of classical dictionary learning that are fundamentally misaligned with the computational model of quantum computing. We identify four key obstacles: greedy and adaptive sparse coding, dense and global dictionary updates, reliance on matrix factorizations, and unavoidable classical output bottlenecks. Each of these alone poses a serious challenge; taken together, they render classical dictionary learning intrinsically incompatible with quantum computation.

\subsection{Greedy and Adaptive Sparse Coding}

Sparse coding is the first and arguably most important step in classical dictionary learning. Given a fixed dictionary, the goal is to represent each data sample as a sparse linear combination of dictionary atoms. In practice, this is almost always solved using greedy algorithms such as Orthogonal Matching Pursuit (OMP) or its variants. These algorithms operate by iteratively selecting atoms that best correlate with the current residual, updating the residual, and repeating until a stopping criterion is met.

This process is inherently adaptive and sequential. Each step depends on the outcome of the previous step, and the control flow branches based on intermediate results. Such behavior is deeply problematic from a quantum computing perspective. Quantum algorithms are designed to exploit parallelism over superpositions, but they do not naturally support adaptive branching without intermediate measurements. Each measurement collapses the quantum state and destroys superposition, eliminating the very mechanism that provides quantum advantage.

Even if one were to attempt a quantum analogue of greedy sparse coding, the algorithm would require repeated measurements and classical feedback at each iteration. This would effectively reduce the computation to a classical loop with quantum subroutines, offering little or no advantage. In short, greedy sparse coding is algorithmically hostile to quantum execution.

\subsection{Convex Relaxations Do Not Resolve the Issue}

One might hope that replacing greedy sparse coding with convex optimization, such as $\ell_1$-minimization, would make the problem more amenable to quantum algorithms. After all, quantum algorithms for solving linear systems and convex problems have been proposed. However, this hope does not withstand closer scrutiny.

First, quantum linear system solvers typically require strong assumptions about data access, most notably the availability of quantum random access memory (QRAM). QRAM is a hypothetical construct that allows coherent access to classical data in superposition. While QRAM is mathematically convenient, it is widely regarded as impractical for near-term quantum hardware and even questionable in the long term.

Second, even if QRAM were available, convex sparse coding methods still produce classical outputs: sparse coefficient vectors that must be explicitly read out. Extracting these coefficients from a quantum state requires either repeated measurements or full state tomography, both of which scale poorly. As a result, any theoretical quantum speedup is overwhelmed by the cost of classical readout.

Finally, convex sparse coding methods often rely on iterative refinement and thresholding operations that are not easily expressed as unitary transformations. These operations again force measurements and classical control, eroding any quantum advantage.

Thus, replacing greedy sparse coding with convex optimization does not fundamentally address the quantum incompatibility of the sparse coding step.

\subsection{Dense Dictionary Updates and Global Operations}

The dictionary update step presents an even more severe obstacle. Classical dictionary learning algorithms such as MOD and K-SVD update the dictionary by solving dense least-squares problems, often involving singular value decomposition (SVD) or QR factorization. These operations are global in nature: updating a single atom typically requires access to the entire dataset and all current sparse codes.

From a quantum perspective, dense matrix factorizations are highly problematic. While quantum algorithms for eigenvalue estimation and matrix inversion exist in theory, they again rely on unrealistic assumptions about data access and coherence. Moreover, these algorithms typically output quantum states rather than explicit classical matrices. Since dictionary learning requires explicit classical dictionaries for subsequent iterations, the quantum output must be fully measured and stored classically.

In practical terms, this means that even a hypothetical quantum-accelerated dictionary update would need to be followed by an expensive classical readout, negating the benefit of the quantum computation. Additionally, the depth and complexity of the required quantum circuits would exceed the capabilities of NISQ-era devices.

\subsection{Lack of Localization and Projection Structure}

Another important issue is the lack of localization in classical dictionary learning. Both sparse coding and dictionary update steps involve dense interactions between atoms and data samples. There is no natural notion of localized updates or simple projections that can be applied independently.

Quantum algorithms are most effective when problems can be decomposed into simple, repeated primitives such as inner products, projections, or rotations in low-dimensional subspaces. Classical DL algorithms, by contrast, rely on dense residual updates and global optimization steps that resist such decomposition. This structural mismatch makes it difficult to map classical DL onto quantum hardware in a natural way.

\subsection{The Classical Output Bottleneck}

Perhaps the most fundamental obstacle is the classical output bottleneck inherent in dictionary learning. The end product of DL is a dictionary matrix that must be explicitly stored and reused in subsequent computations. Unlike classification or sampling tasks, where the output may remain implicit or probabilistic, dictionary learning demands full classical access to the learned parameters.

Quantum computation excels when the result of a computation can be kept in quantum form or when only limited information needs to be extracted. Dictionary learning violates this assumption. Regardless of how efficiently intermediate steps are performed, the final dictionary must be fully materialized classically. This requirement alone places a hard limit on the potential for quantum advantage.

\subsection{Summary: A Structural Mismatch}

Taken together, these issues reveal a clear pattern. Classical dictionary learning algorithms are built around greedy, adaptive procedures, dense global operations, and explicit classical outputs. These characteristics are not accidental; they are central to the effectiveness of DL in classical settings. Unfortunately, they are also precisely the characteristics that make an algorithm unsuitable for quantum execution.

The repeated failure to develop practical quantum dictionary learning algorithms should therefore be understood not as a lack of creativity, but as evidence of a fundamental structural mismatch. Attempting to directly quantize classical DL is unlikely to succeed, regardless of incremental algorithmic modifications.

This observation motivates a shift in perspective. Rather than asking how to accelerate classical dictionary learning using quantum computation, we should ask how dictionary learning itself can be reformulated to align with the strengths and limitations of quantum hardware. In the next section, we show that doubly sparse dictionary learning provides exactly such a reformulation.

\section{Doubly Sparse Dictionary Learning as a Structural Resolution}

The analysis in the previous section leads to a clear conclusion: classical dictionary learning fails to translate to quantum computation because of its structural properties, not because of inefficient implementations. If dictionary learning is to be made compatible with quantum computation, its structure must change. Doubly sparse dictionary learning (DSDL) provides exactly such a structural modification.

In DSDL, the dictionary is no longer learned as an unconstrained dense matrix. Instead, it is expressed as a sparse combination of atoms from a known and fixed basis. Concretely, the dictionary is modeled as
\[
D = \Phi A,
\]
where $\Phi$ is a known basis (such as wavelets, Fourier atoms, or other structured transforms), and $A$ is a sparse coefficient matrix. Both the representation coefficients and the dictionary parameters are sparse, hence the term \emph{doubly sparse}.

This formulation was originally introduced in the classical signal processing literature as a way to reduce the number of free parameters, improve interpretability, and incorporate domain knowledge. In the present context, however, its most important role is structural: it fundamentally alters the algorithmic geometry of dictionary learning in a way that aligns with quantum computation.

\subsection{Why Constraining the Dictionary Matters}

In classical dictionary learning, the dictionary is a dense object with no predefined structure. Each update to the dictionary potentially affects every atom and every signal. This global coupling is what necessitates dense least-squares solves and matrix factorizations such as SVD.

By contrast, in DSDL the dictionary is built from a known basis. The learning problem is no longer about discovering arbitrary directions in the ambient space, but about selecting and combining elements from a fixed, structured set. This has several immediate consequences.

First, the number of degrees of freedom is drastically reduced. Rather than learning an entire dense matrix, the algorithm only needs to learn the sparse coefficients in $A$. This reduction not only simplifies the learning problem but also makes it more amenable to iterative, projection-based methods.

Second, dictionary updates become localized in the basis domain. Updating a single atom of $D$ corresponds to updating a sparse vector in the basis $\Phi$. This localization eliminates the need for dense global updates and enables simple linear projections to drive learning.

Third, the known basis $\Phi$ can be chosen to reflect prior knowledge about the data, such as multiscale structure or frequency localization. This choice is entirely classical and does not impose additional quantum requirements.

\subsection{Reframing Sparse Coding as Projection}

The sparse coding step also undergoes a qualitative transformation in the DSDL framework. Instead of solving a sparse approximation problem with respect to an arbitrary dense dictionary, sparse coding becomes the task of approximating data using a product of a known basis and a sparse coefficient matrix.

This problem can be expressed as a sequence of linear projection steps rather than as a greedy or convex optimization procedure. Iterative projection methods, such as the Kaczmarz algorithm, are well suited to this task. These methods operate by repeatedly projecting the current estimate onto solution spaces defined by individual equations or constraints.

Projection-based methods have several properties that are highly desirable from a quantum computing perspective:
\begin{itemize}
    \item They rely primarily on inner products and vector updates.
    \item They do not require adaptive branching or support selection.
    \item They can be implemented as simple, repeated primitives.
\end{itemize}

Early stopping of these iterations naturally induces sparsity by preventing overfitting, playing a role similar to regularization or thresholding but without explicit non-linear operations.

\subsection{Dictionary Update Without Matrix Factorization}

Perhaps the most significant advantage of DSDL is that dictionary updates no longer require matrix factorizations. In classical DL, updating the dictionary often involves solving a dense least-squares problem over all data samples, which in turn requires SVD or QR decomposition.

In DSDL, dictionary update reduces to estimating sparse coefficients in the known basis $\Phi$. This task is again a linear projection problem and can be solved using the same class of iterative methods as sparse coding. The symmetry between the two steps simplifies the algorithmic structure and avoids the need for fundamentally different solvers.

This symmetry is crucial for quantum compatibility. A single projection-based solver can be reused throughout the algorithm, with quantum computation appearing only in the estimation of inner products. No step requires a quantum analogue of matrix factorization, which remains one of the most challenging operations to implement quantumly.

\subsection{Alignment with Quantum Computational Primitives}

Quantum algorithms are most naturally expressed in terms of linear transformations, inner products, and projections in Hilbert spaces. The DSDL formulation aligns closely with this paradigm. Both sparse coding and dictionary update can be expressed using repeated projections driven by inner product computations.

Importantly, the algorithm does not require the quantum computer to store or manipulate the entire dataset in superposition. Data remains classical, and only small vectors are encoded into quantum states for the purpose of estimating overlaps. This hybrid approach avoids strong assumptions such as QRAM and keeps quantum circuits shallow.

From this perspective, quantum computation is not used to replace the entire learning algorithm, but to accelerate or simplify a specific primitive that appears repeatedly: inner product estimation. This limited but realistic role for quantum computation is consistent with the capabilities of near-term devices.

\subsection{Comparison with Classical Dictionary Learning}

The structural differences between classical dictionary learning and DSDL can be summarized along several dimensions:

\begin{itemize}
    \item \textbf{Parameterization}: Classical DL learns a dense dictionary, while DSDL learns sparse coefficients in a fixed basis.
    \item \textbf{Sparse coding}: Classical DL relies on greedy or convex solvers; DSDL uses projection-based iterative methods.
    \item \textbf{Dictionary update}: Classical DL uses dense factorizations; DSDL uses linear projections.
    \item \textbf{Quantum role}: Classical DL has no natural quantum primitive; DSDL relies on inner products, which are quantum-friendly.
\end{itemize}

These differences are not superficial. They represent a shift from global, adaptive optimization to local, iterative refinement. This shift is what makes DSDL structurally compatible with quantum computation.

\subsection{A Change in Perspective}

It is important to emphasize that DSDL is not merely a constrained version of classical dictionary learning. It represents a different way of thinking about representation learning. Instead of discovering arbitrary directions in data space, DSDL focuses on selecting and combining elements from a structured basis.

From a classical perspective, this may appear restrictive. From a quantum perspective, however, it is precisely this structure that enables compatibility. The restriction is not a weakness but a design choice that aligns the learning problem with the computational model.

This change in perspective suggests a broader lesson for quantum machine learning. Rather than attempting to quantize classical algorithms directly, it may be more fruitful to rethink the underlying models and objectives so that they naturally align with quantum primitives.

In the next section, we build on this structural insight to present a concrete hybrid quantum--classical algorithm for doubly sparse dictionary learning based on randomized projection methods.

\section{Algorithmic Framework and Quantum-Compatible Design}

Having established that doubly sparse dictionary learning (DSDL) provides a structurally compatible formulation for quantum computation, we now describe the algorithmic framework used in this work. Our goal is not to present a mathematically optimal solver, but to design an algorithm that respects the constraints of near-term quantum hardware while remaining faithful to the DSDL model.

The guiding principle is simplicity: every component of the algorithm should reduce to repeated applications of a small number of primitives. In particular, we aim to express both sparse coding and dictionary update in terms of linear projections driven by inner products. Quantum computation is used only where it is most natural and realistic: estimating inner products between vectors.

\subsection{Design Principles}

The algorithm is guided by the following design principles:

\begin{itemize}
    \item \textbf{Projection-based updates}: All optimization steps are formulated as projections rather than exact solves or greedy selection.
    \item \textbf{Shared solver structure}: Sparse coding and dictionary update use the same iterative mechanism.
    \item \textbf{Minimal quantum role}: Quantum computation is limited to inner product estimation.
    \item \textbf{Hybrid execution}: Data storage, control flow, and parameter updates remain classical.
    \item \textbf{Stability over optimality}: Regularization and normalization are prioritized over exact convergence.
\end{itemize}

These principles reflect a deliberate choice to prioritize implementability and robustness over asymptotic guarantees.

\subsection{Randomized Kaczmarz as a Core Primitive}

At the heart of the algorithm is the randomized Kaczmarz method. The classical Kaczmarz algorithm solves linear systems by iteratively projecting the current estimate onto the solution space of a randomly selected equation. The method is simple, memory-efficient, and naturally expressed in terms of inner products and vector updates.

Randomized Kaczmarz has several properties that make it attractive in the present context. First, each iteration requires only a single row of the system matrix, avoiding dense global operations. Second, the update rule involves only an inner product and a scaled vector addition. Third, the algorithm admits early stopping, which provides a natural form of implicit regularization.

In our framework, both sparse coding and dictionary update are reduced to solving linear systems of the form
\[
Mx \approx y,
\]
where $M$ is either the effective dictionary $D = \Phi A$ or the known basis $\Phi$, depending on the step. The same randomized projection mechanism can therefore be reused throughout the algorithm.

\subsection{Sparse Coding via Early-Stopped Projections}

In the sparse coding step, each data sample is approximated using the current dictionary $D = \Phi A$. Rather than seeking an exact sparse solution, we apply a fixed number of randomized Kaczmarz iterations starting from a zero initialization.

Early stopping plays a crucial role here. By limiting the number of iterations, we prevent the solution from fully converging to a dense least-squares estimate. The resulting approximation is typically sparse or compressible, even without explicit sparsity constraints. This approach avoids greedy atom selection and non-linear thresholding, both of which are problematic for quantum implementation.

From a quantum perspective, this step is well suited to hybrid execution. The classical controller selects rows and updates coefficients, while the quantum device is invoked to estimate inner products needed for the projection updates.

\subsection{Dictionary Update in a Known Basis}

The dictionary update step follows a similar pattern. Instead of updating the dictionary directly, we update the sparse coefficient matrix $A$ in the known basis $\Phi$. For each column of $A$, we form a residual that captures the contribution of the corresponding dictionary atom and apply randomized Kaczmarz iterations to estimate an update.

This step avoids dense matrix factorizations entirely. Each update depends only on projections in the basis $\Phi$, which is fixed and known. The update is followed by normalization to prevent scale ambiguity and ensure numerical stability.

The symmetry between sparse coding and dictionary update is intentional. Both steps rely on the same primitive operations and differ only in the choice of system matrix. This uniformity simplifies both implementation and analysis.

\subsection{Ridge Stabilization and Normalization}

Projection-based methods are known to be sensitive to noise and ill-conditioning. To address this, we incorporate ridge stabilization into the Kaczmarz updates. Specifically, a small regularization term is added to the denominator of each update, preventing excessive amplification of noise.

Normalization of dictionary atoms is also essential. Without normalization, the alternating updates can lead to unbounded growth in one component and corresponding shrinkage in another. Normalization enforces a consistent scale and removes this degree of freedom.

These stabilization mechanisms are classical in nature and do not interfere with quantum compatibility. On the contrary, they are essential for ensuring that the algorithm remains stable under approximate inner product estimation.

\subsection{Quantum Inner Product Estimation}

The only step that explicitly involves quantum computation is inner product estimation. In each projection update, the algorithm requires the inner product between two vectors. This operation can be implemented on a quantum device by encoding the vectors as quantum states and estimating their overlap.

In our implementation, we use Qiskit to construct quantum states corresponding to normalized vectors and compute their overlaps using standard statevector or circuit-based primitives. The circuits involved are shallow and do not require entanglement across large numbers of qubits.

Crucially, the algorithm does not assume that all data is stored in quantum form. Vectors are loaded into quantum states only when needed, and the resulting inner product estimates are immediately returned to the classical controller. This approach avoids QRAM assumptions and keeps the quantum component lightweight.

\subsection{Hybrid Control Flow}

The overall algorithm follows a classical control loop. At each iteration, the classical controller orchestrates sparse coding and dictionary update steps, invoking quantum subroutines only for inner product estimation. This hybrid architecture reflects a realistic deployment scenario for near-term quantum devices.

The quantum component is stateless and ephemeral: it computes an inner product and returns a scalar. All stateful learning parameters reside in classical memory. This design avoids coherence requirements across iterations and simplifies error handling.

\subsection{Implementation and Reproducibility}

A complete reference implementation of the algorithm is provided in an open-source repository:
\begin{center}
\url{https://github.com/AngshulMajumdar/quantum-dsdl-kaczmarz}
\end{center}

The implementation is intentionally minimal and flat, reflecting the algorithmic simplicity of the approach. It serves as a proof of concept rather than a performance-optimized library. The emphasis is on clarity, reproducibility, and structural alignment with quantum computation.

In the next section, we discuss practical observations regarding the behavior of the algorithm, including stability, convergence characteristics, and limitations arising from stochastic projection updates.

\section{Practical Notes, Stability, and Limitations}

The algorithmic framework described in the previous section is intentionally simple and projection-based. As a result, its behavior differs in important ways from classical dictionary learning algorithms that rely on exact or near-exact optimization at each step. In this section, we summarize practical observations that arise when the algorithm is implemented and executed, and we clarify the scope and limitations of the approach.

\subsection{Non-Monotonic Objective Behavior}

A key practical observation is that the reconstruction error is not guaranteed to decrease monotonically across outer iterations. This behavior is a direct consequence of two design choices: the use of randomized projection methods and early stopping. Each sparse coding and dictionary update step solves its corresponding subproblem only approximately, and the stochastic nature of row sampling introduces variance.

Empirically, we observe a consistent pattern across runs: the reconstruction error decreases substantially during the initial iterations and then enters a regime of bounded oscillations. This behavior is consistent with the interpretation of the algorithm as a stochastic, projection-based alternating method rather than a deterministic optimizer. Importantly, the iterates remain stable and do not diverge when ridge stabilization and normalization are applied.

From the perspective of this work, monotonic descent is neither required nor desirable. Enforcing monotonicity would require exact solves, dense factorizations, or explicit sparsity projections, all of which undermine quantum compatibility. The observed behavior reflects a bias--variance tradeoff inherent to early-stopped iterative methods.

\subsection{Implicit Sparsity via Early Stopping}

Unlike classical dictionary learning algorithms, which explicitly enforce sparsity using greedy selection or convex penalties, our approach relies on early stopping to induce sparsity implicitly. This mechanism is well understood in iterative methods and has a natural interpretation as regularization.

Early stopping prevents the solution from converging to a dense least-squares estimate. Instead, the algorithm captures the most significant components of the representation before variance dominates. This approach avoids non-linear thresholding operations and support recovery, both of which are problematic in quantum settings.

While implicit sparsity may not achieve the same level of sparsity as aggressive greedy methods, it offers a principled and quantum-compatible alternative that aligns with the projection-based nature of the algorithm.

\subsection{Role of Ridge Stabilization and Normalization}

Ridge stabilization and normalization play a crucial role in ensuring numerical stability. Without ridge regularization, projection-based updates can amplify noise, particularly when rows of the system matrix have small norms. The addition of a small regularization term mitigates this effect and ensures bounded updates.

Normalization of dictionary atoms is equally important. In alternating minimization schemes, scale ambiguity between dictionary atoms and coefficients can lead to unbounded growth in one component and corresponding shrinkage in another. Normalization removes this degree of freedom and stabilizes learning.

These stabilization techniques are classical, lightweight, and compatible with approximate inner product estimation. They do not introduce additional quantum requirements and are essential for practical execution.

\subsection{Computational Considerations}

The algorithm is computationally lightweight on the classical side, relying primarily on vector updates and random sampling. The quantum component is limited to inner product estimation between normalized vectors. In the reference implementation, these inner products are computed using Qiskit statevector primitives, but the same logic applies to circuit-based overlap estimation on real hardware.

Crucially, the algorithm does not assume the availability of QRAM or the ability to store large datasets in quantum memory. All data remains classical, and quantum computation is invoked only for small, ephemeral tasks. This design reflects a realistic deployment model for near-term quantum devices.

\subsection{Limitations}

It is important to be explicit about what this approach does not claim.

First, the algorithm does not provide exponential or even asymptotic quantum speedups for dictionary learning. Any potential advantage arises from constant-factor improvements in specific primitives and from structural compatibility, not from asymptotic complexity reductions.

Second, the learned dictionaries are constrained by the choice of basis $\Phi$. While this constraint is essential for quantum compatibility, it may limit expressiveness compared to unconstrained classical dictionaries. This tradeoff is inherent and should be viewed as a design choice rather than a deficiency.

Third, the algorithm is not intended as a drop-in replacement for highly optimized classical dictionary learning methods in performance-critical applications. Its value lies in clarifying the structural requirements for quantum compatibility and demonstrating a viable hybrid approach.

\section{Broader Implications for Quantum Machine Learning}

The perspective developed in this paper has implications beyond dictionary learning. It suggests a general principle for quantum machine learning: algorithms that rely on greedy selection, dense global operations, and explicit classical outputs are unlikely to benefit from quantum acceleration in practice.

Conversely, models and algorithms that can be expressed using repeated linear projections, local updates, and approximate primitives are more naturally aligned with quantum computation. The success of the present approach hinges not on sophisticated quantum subroutines, but on reformulating the learning problem itself.

This observation encourages a shift in focus within quantum machine learning research. Rather than asking how to quantize classical algorithms, we should ask how learning models can be redesigned to respect the strengths and limitations of quantum hardware. Doubly sparse dictionary learning provides a concrete example of how such redesign can lead to quantum-compatible algorithms without unrealistic assumptions.

\section{Conclusion}

We have argued that the absence of practical quantum dictionary learning algorithms reflects a fundamental structural mismatch between classical dictionary learning and quantum computation. Greedy sparse coding, dense dictionary updates, reliance on matrix factorizations, and classical output requirements collectively prevent meaningful quantum realization of classical DL.

By adopting a doubly sparse model, we remove these obstacles and arrive at a formulation that is naturally compatible with quantum computation. Using projection-based randomized Kaczmarz iterations and quantum inner product estimation, we present a simple hybrid algorithm that avoids greedy solvers, matrix decompositions, and QRAM assumptions.

Our contribution is not a fully quantum dictionary learning algorithm, nor a claim of quantum speedup. Instead, it is a principled reformulation that clarifies when dictionary learning can be aligned with quantum computation. We believe this perspective provides a useful template for future work at the intersection of representation learning and quantum computing.

A complete reference implementation accompanying this paper is available at:
\begin{center}
\url{https://github.com/AngshulMajumdar/quantum-dsdl-kaczmarz}
\end{center}

\end{document}